\documentstyle[prl,aps,epsf,psfig]{revtex}
\begin{document}
\twocolumn[\hsize\textwidth\columnwidth\hsize
           \csname @twocolumnfalse\endcsname
\title{Duration and non-locality of a nucleon-nucleon collision}
\author{Klaus Morawetz}
\address{Fachbereich Physik, University Rostock, D-18055 Rostock,
Germany}
\author{Pavel Lipavsk\'y and V\'aclav \v Spi\v cka}
\address{Institute of Physics, Academy of Sciences, Cukrovarnick\'a 10,
16200 Praha 6, Czech Republic}
\author{Nai-Hang Kwong}
\address{Physics Department,
University of Arizona,Tucson,Arizona 85721}
\maketitle
\begin{abstract}
For a set of realistic nucleon-nucleon potentials we evaluate microscopic
parameters of binary collisions: a time duration of the scattering state, 
a mean distance and a rotation of nucleons during a collision. These 
parameters enter the kinetic equation as non-instantaneous and non-local 
corrections of the scattering integral, i.e., they can be experimentally
tested. Being proportional to off-shell derivatives of the scattering 
T-matrix, non-instantaneous and non-local corrections make it possible to
compare the off-shell behavior of different potentials in a vicinity of
the energy shell. The Bonn one-Boson-exchange (A-C) and Paris potentials 
are found to yield very close results, while the separable Paris potential 
differs.
\end{abstract}
\vskip2pc]

\section{Introduction}
Most of realistic studies of heavy ion reactions deal with local and
instantaneous binary collisions as they appear in the Boltzmann, or the
Boltzmann-Uehling-Uhlenbeck (BUU), equation. The importance of the
non-local treatment of collisions has been first pointed out by Halbert
\cite{H81} and Malfliet \cite{M84} and more recently also studied by 
Kortemeyer, Daffin and Bauer \cite{KDB96}. In these studies, nucleons 
are treated as hard spheres, i.e., each collision is instantaneous and
a distance of nucleons at the instant of collision equals the diameter
of the scattering cross section. This is a questionable approximation,
a further progress in these studies hence requires to evaluate the 
non-local corrections from a microscopic picture.

The first step in this direction was done by Danielewicz and Pratt 
\cite{DP96}. They have implemented the Beth-Uhlenbeck approach 
supplemented with the Wigner concept of the collision delay to evaluate 
virial corrections to the pressure of the nuclear matter. At a first 
glance, the correction discussed by Danielewicz and Pratt is quite 
different from \cite{H81,M84,KDB96}. Indeed, the hard-sphere collisions
are instantaneous and non-local while in the Wigner approach all 
collisions are local but non-instantaneous. A comparison of both 
concepts is not so straightforward, however. The Wigner collision 
delay does not describe a duration of collision but provides the best 
one-parameter fit of the scattered wave in the asymptotic region. 
In particular, the Wigner collision delay for the collision of hard 
spheres is non-zero because a particle reflected by a hard-sphere 
potential arrives at the asymptotic region sooner then it would do in 
a reference point-like (local) scattering event.  In this way, the 
Wigner collision delay represents a mixture of the collision duration 
and a certain approximation of non-local corrections. The quantitative 
comparison of both approaches is possible. Numerical values of the 
Wigner collision delay, found by Danielewicz and Pratt from the 
scattering T-matrix, differ from the hard-sphere estimate.
 
A picture of collision, which is non-local and non-instantaneous in 
the same time, has resulted from the Green's function approach to 
the kinetic equation \cite{SLM96}. As in the Wigner concept, there is 
a characteristic time $\Delta_t$ determined from the scattering T-matrix. 
The time $\Delta_t$ vanishes in the limit of hard spheres what shows 
that it can be interpreted as a duration of collision.\footnote{
In papers \cite{SLM96} and \cite{MLSCN98} the terms {\em collision 
duration} and {\em collision delay} are treated as equivalent. Here we 
use the {\em collision delay} exclusively for the Wigner concept while 
the {\em collision duration} for the concept of \cite{SLM96}.}
The non-local corrections, also determined from the T-matrix, are treated 
separately. In the limit of hard spheres they yield the displacement 
which equals the diameter of the hard sphere. The non-local and 
non-instantaneous theory furnishes us with a link between both approaches 
discussed above. If one neglects collisions on the time scale of the 
collision duration, the collision found in \cite{SLM96} can be recast 
into an instantaneous and non-local event. This approximation was used 
in \cite{MLSCN98} as a convenient tool to incorporate the collision 
duration together with non-local corrections into realistic simulations 
of heavy ion reactions. Similarly, the Wigner concept is achieved, if 
one neglects the angular momentum of colliding particles. 

In this paper we discuss numerical values of the collision duration, the
hard-sphere-type displacement of particles, and the displacement related
to the angular momentum. Following \cite{SLM96}, we evaluate these
parameters of non-local and non-instantaneous corrections from the
scattering T-matrix. Presented results show that the Wigner collision
delay is a good approximation at the low energy region where the model 
of hard spheres is not adequate. For higher energies, the angular and 
energy dependencies of the non-local and non-instantaneous corrections
are too complicated to be captured by a simple model.

Numerical values of non-local and non-instantaneous corrections depend 
on the interaction potential one employs. Although this dependence is not
dramatic, it is stronger than the dependence of the differential cross 
section. Since values of corrections can, at least in principle, be 
inferred from heavy ion reactions, they promise a supplementary test of 
interaction potentials. 

\section{Parameters of non-local collisions}
A picture of the non-local collision is outlined in Fig.~\ref{soft}. It 
presents an event in which the particle $a$ of the momentum $k$ scatters 
with the particle $b$ of the momentum $p$ loosing the momentum $q$, i.e.,
$a$ ends with $k-q$ and $b$ ends with $p+q$. At the beginning of the 
collision, these particles are displaced from each other by $\Delta_2$. 
Placing the particle $a$ into the initial of coordinate, we associate 
$\Delta_2$ with the initial position of particle $b$. During the collision 
both particles move, accordingly, at the end of the collision the particle 
$a$ has the position $\Delta_3$ and the particle $b$ has the position 
$\Delta_4$. 

\subsection{Elementary displacements}
\begin{figure}
\begin{minipage}{7cm}
\parbox[h]{7cm}{
  \psfig{figure=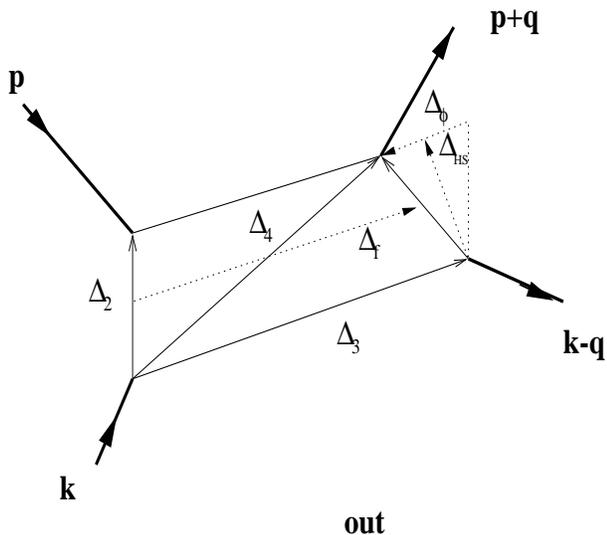,width=8cm,height=7cm}}
\end{minipage}
\vspace{2ex} 
\caption{The non-local binary collision.}
\label{soft}
\end{figure}

The displacements $\Delta_{2,3,4}$ can be expressed in terms of 
quantities with a more transparent physical meaning. During the 
collision lasting for $\Delta_t$, the center of mass of the colliding 
pair flies over a distance
\begin{equation}
\Delta_{\rm f}={k+p\over 2m}\Delta_t,
\label{e1}
\end{equation}
where we assume that the particles have equal masses, $m$. Since the 
colliding pair has a non-zero angular momentum, during $\Delta_t$ they 
rotate around each other so that their relative displacement changes by 
$2\Delta_\phi$. The last independent vector can be selected from the
symmetry. The flight $\Delta_{\rm f}$ and the rotation $\Delta_\phi$ 
reverse their orientations under time inversion. It is advantageous to 
complement them with the mean of initial and final distance, 
$\Delta_{\rm HS}$ seen in Fig.~\ref{soft}, which is invariant under 
time inversion. The positions of the particles are linked with the 
elementary displacements by relations evident from Fig.~\ref{soft}
\begin{eqnarray}
2\Delta_\phi&=&\Delta_4-\Delta_3-\Delta_2\nonumber\\
2\Delta_{\rm HS}&=&\Delta_4-\Delta_3+\Delta_2 \nonumber\\
2\Delta_{\rm f}&=&\Delta_4+\Delta_3-\Delta_2.
\label{para}
\end{eqnarray}

The direction of $\Delta_{\rm f}$ is identical to the centre of mass
velocity, see (\ref{e1}). For central forces, both remaining displacements 
lay in the collision plane. The direction of $\Delta_{\rm HS}$ is then 
identical to the direction of the transferred momentum $q$. We note that
in the limit of hard spheres, $|\Delta_{\rm HS}|$ equals the sphere 
diameter while the other $\Delta$'s vanish. The rotational displacement 
$\Delta_\phi$ is orthogonal to $\Delta_{\rm HS}$ having the direction 
of $k-p-q$. Accordingly, only three scalar parameters, $\Delta_t$, 
$|\Delta_{\rm HS}|$ and $|\Delta_\phi|$, are sufficient to characterize
the non-instantaneous and non-local features of the collision.

\subsection{Collision duration versus delay}
As pointed above, the collision duration differs from the Wigner 
collision delay. It is profitable to outline why the duration and not the
delay is more suitable to describe properties of nucleon-nucleon 
collisions.

One can observe the collision from two different aspects. First, in the
asymptotic region the non-local and non-instantaneous corrections 
are aimed to provide the exact trajectory of the scattered wave packet.
From this asymptotic point of view, one of the three above parameters 
is redundant. Indeed, the displacement in the direction of the outgoing 
trajectory can be recast into the modified collision duration, which 
is just the Wigner collision delay. Alternatively, one can extrapolate 
the incoming and outgoing trajectories so that the collision duration 
is suppressed and the collision is effectively instantaneous, see 
\cite{MLSCN98}. 

Second, one can observe a dense system and try to measure what share of 
nucleons just undergo a collision. This share is given by the so called 
correlated density \cite{SRS90}. The correlated density depends on the 
collision duration but not on the displacements, see \cite{SLM96}. To 
cover both aspects, all the three parameters are necessary and the 
collision duration, not the Wigner collision delay, has to be used.

\subsection{Relation to T-matrix}
To evaluate the collision duration and the displacements one needs the
retarded T-matrix $T(\Omega,k,p,q)$, as a function of the incoming and 
transferred momenta and an independent energy $\Omega$. The $\Delta$'s
given by derivatives of the scattering phase shift ${\rm Im}\ln T$.
According to \cite{SLM96} the collision duration reads
\begin{equation}
\Delta_t={\rm Im}{1\over T}{\partial T\over\partial\Omega}.
\label{delt}
\end{equation}
Note that one has to evaluate the T-matrix out of the energy shell, 
although only the on-shell value, $\Omega={k^2\over 2m}+{p^2\over 2m}$, 
of the collision duration appears as the non-instantaneous correction in 
the kinetic equation. The on-shell condition is used after taking the
derivative. From (\ref{para}) and $\Delta_{2,3,4}$ found in \cite{SLM96} 
one obtains the hard-sphere and rotational displacements 
\begin{eqnarray}
\Delta_{\rm HS}&=&{\rm Im}{1\over 2T}\left({\partial T\over\partial p}-
{\partial T\over\partial k}-2{\partial T\over\partial q}\right),
\nonumber\\
\Delta_\phi&=&{\rm Im}{1\over 2T}\left({\partial T\over\partial k}-
{\partial T\over\partial p}\right).
\label{8}
\end{eqnarray}
Again, $\Omega$ is an independent variable and the on-shell value is 
taken after derivatives.

\section{Free-space collisions}
Our aim is to evaluate the collision duration $\Delta_t$, the 
hard-sphere displacement $\Delta_{\rm HS}$ and the rotational 
displacement $\Delta_\phi$ for a collision of two nucleons in the free 
space. The translational, rotational and time-reversal symmetries
make it possible to express all these quantities in terms of the 
decomposition into partial waves.

\subsection{Symmetries}
Formulas (\ref{delt}) and (\ref{8}) are expressed in terms of momenta,
$(k,p,q)$, commonly met in the scattering integral of the kinetic 
equation. For the numerical evaluation, it is advantageous to rearrange 
them into the barycentric framework where one can more conveniently 
handle symmetries of collisions. To this end we write the T-matrix as 
a function of total momentum $K$, initial relative momentum $\kappa$ 
and final relative momentum $\kappa_f$,
\begin{eqnarray}
K&=&k+p
\nonumber\\
\kappa &=&{1\over 2}(k-p)
\nonumber\\
\kappa_f &=&{1\over 2}(k-p)-q.
\label{barm}
\end{eqnarray}
In this representation the hard-sphere and rotational displacements 
read
\begin{eqnarray}
\Delta_{\rm HS}&=&-{\rm Im}{1\over 2T}\left({\partial T
\over\partial\kappa}-{\partial T\over\partial\kappa_f}\right),
\nonumber\\
\Delta_{\phi}&=&{\rm Im}{1\over 2T}\left({\partial T
\over\partial\kappa}+{\partial T\over\partial\kappa_f}\right),
\label{barda}
\end{eqnarray}
and the collision duration is defined by (\ref{delt}).

The nucleon-nucleon interaction includes the tensor forces, therefore
the scattering has to be treated with the theory for non-central forces. 
To avoid this complication, in the present treatment we will focus on 
the trace element of the decomposition of the T-matrix into singlet and 
triplet channels. This element obeys the same symmetries as the T-matrix 
of scattering on central forces. To avoid redundant symbols, from now
on the T-matrix means the approximation by its trace component. This 
approximation is specified below, see (\ref{tmsch}).

The trace element of the T-matrix depends only on the relative initial 
and final kinetic energies measured by $|\kappa|$ and $|\kappa_f|$, 
respectively, and on the deflection angle $\theta$, given by 
$\cos\theta={\kappa\kappa_f\over|\kappa||\kappa_f|}$,
\begin{equation}
T\equiv T\left(\Omega-{K^2\over 4m},\cos\theta,|\kappa|,|\kappa_f|
\right).
\label{bart}
\end{equation}
The independent energy $\Omega$ is reduced by the center-of-mass kinetic 
energy ${K^2\over 4m}$ due to the transformation into the barycentric
framework.

Apparently, the T-matrix (\ref{bart}) depends on four scalar arguments, 
therefore there can be at maximum four independent derivatives. This 
number is reduced to three by the space-reversal and the time-reversal 
symmetries which require \cite{GW64}
\begin{equation}
T(\Omega,\cos\theta,|\kappa|,|\kappa_{\rm f}|)=
T(\Omega,\cos\theta,|\kappa_{\rm f}|,|\kappa|).
\label{mc33}
\end{equation}
The displacements are evaluated on the energy shell, where (after taking
derivatives) the amplitudes of the initial and the final momenta equal
each other, $|\kappa_{\rm f}|=|\kappa|$. From the symmetry (\ref{mc33})
then follows
\begin{eqnarray}
\left.{\partial T(|\kappa|,|\kappa_{\rm f}|)\over\partial|\kappa_{\rm f}|}
\right|_{|\kappa_{\rm f}|=|\kappa|}&=&
\left.{\partial T(|\kappa|,|\kappa_{\rm f}|)\over\partial|\kappa|}
\right|_{|\kappa_{\rm f}|=|\kappa|}
\nonumber\\
&=&{1\over 2}{\partial T(|\kappa|,|\kappa|)\over\partial|\kappa|}
\equiv{1\over 2}{\partial T(|\kappa|)\over\partial|\kappa|}.
\label{mc35}
\end{eqnarray}
Accordingly, only three of the derivatives are independent, in agreement 
with three parameters needed to describe the non-instantaneous and 
non-local features of collisions. From now on we use the last form in 
the second line. 

The symmetry determines the direction of displacements. On the energy
shell the vector derivatives simplify as
\begin{eqnarray}
&&\left.\left({\partial\over\partial\kappa}\pm{\partial
\over\partial\kappa_{\rm f}}\right)T(\cos\theta,|\kappa|,
|\kappa_{\rm f}|)\right|_{|\kappa_{\rm f}|=|\kappa|}
\nonumber\\
&&=(\kappa\pm\kappa_{\rm f})\left({1\over 2|\kappa|}{\partial
\over\partial|\kappa|}-{\cos\theta\mp 1\over|\kappa|^2}{\partial
\over\partial\cos\theta}\right)T(\cos\theta,|\kappa|).
\nonumber\\
\label{derexp}
\end{eqnarray}
From (\ref{barda})  and (\ref{derexp}) one can see that the 
displacements, 
\begin{eqnarray}
\Delta_{\rm HS}&=&\overline{\Delta_{\rm HS}}
{\kappa-\kappa_f\over|\kappa-\kappa_f|},
\nonumber\\
\Delta_{\phi}&=&\overline{\Delta_\phi}
{\kappa+\kappa_f\over|\kappa+\kappa_f|}.
\label{bard}
\end{eqnarray}
follow directions expected from the symmetries. The hard-sphere 
displacement is parallel to the transferred momentum $q=\kappa-\kappa_f$.
The rotational displacement stays in the collision plane and is 
orthogonal to the transferred momentum. The orthogonality follows from 
the energy conservation, $(\kappa+\kappa_f)(\kappa-\kappa_f)=\kappa^2-
\kappa_f^2=0$.

The displacement lengths, $\overline{\Delta_\phi}$ and $\overline
{\Delta_{\rm HS}}$, are defined by (\ref{bard}). Their values are easily 
traced down from (\ref{barda}) and (\ref{derexp}). Below we provide 
their explicit forms. We call them lengths to emphasize that they are 
scalars, they can, however, assume both positive and negative values. 

It is noteworthy that for the zero-angle scattering, $\theta=0^o$ or
$\kappa_f=\kappa$, the transfer momentum is zero so that its direction
is not defined. The hard-sphere displacement thus cannot be constructed.
This is a peculiarity of the quantum picture of collisions within which
the collision plane is determined by the initial and final momenta.
Within the classical picture, the collision plane can be determined from
the initial momentum and the impact parameter. According to
Fig.~\ref{soft}, the hard-sphere displacement then equals to the impact
parameter and is perpendicular to the initial momentum. Similarly, for 
back scattering, $\theta=180^o$ or $\kappa_f=-\kappa$, the
direction of the rotational displacement is not defined within the quantum
picture. As we will see below, this problem is only academic since both
displacements vanish in the questionable cases.

\subsection{Decomposition in partial waves}
It remains to evaluate the collision delay and amplitudes of the
displacements. To this end we use the decomposition of the spin averaged 
T-matrix into partial waves
\begin{equation}
T(\Omega,\cos\theta,|\kappa|)={1\over 16\pi}\sum\limits_l
P_l(\cos\theta)T_l(\Omega,|\kappa|),
\label{tma}
\end{equation}
where $P_l$ is a Legendre polynomial. The coefficients $T_l$ are on the
shell $|\kappa_f|=|\kappa|$. They are linked to the usual channel
partial T-matrices \cite{GW64}
\begin{equation}
T_l=\sum\limits_{I,S=0,1}\sum\limits_{J=|S-l|}^{S+l}
(2J+1)(2I+1)T_{l,l}^{S,I,J}
\label{tmsch}
\end{equation}
with the total spin $S$, isospin $I$ and the total angular momentum $J$.
Due to summation over projection components $m_l, m_s$ only diagonal
elements in $l$ contribute. 

Substitution of decomposition (\ref{tma}) into (\ref{delt}) and
(\ref{barda}) with (\ref{derexp}) yields
\begin{eqnarray}
\Delta_t&=&{\rm Im}\left[{1\over T}\sum\limits_lP_l
{\partial T_l\over\partial\Omega}\right]
\nonumber\\
\overline{\Delta_{\rm HS}}&=&{{\sin\theta/2}\over|\kappa|}\ {\rm Im}\left[
{1\over T}\sum\limits_lP'_lT_l(1+\cos\theta)-|\kappa|P_lT'_l\right]
\nonumber\\
\overline{\Delta_\phi}&=&{\cos\theta/2\over|\kappa|}\ {\rm Im}\left[
{1\over T}\sum\limits_lP'_lT_l(1-\cos\theta)+|\kappa|P_lT'_l\right],
\nonumber\\
\label{tmexp}
\end{eqnarray}
where $P_l'(z)={\partial\over\partial z}P_l$ is the derivative of the
Legendre polynomial and $T'_l={1\over 2}{\partial\over\partial|\kappa|}
T_l(|\kappa|)$. 

Quantum formulas (\ref{tmexp}) show that the hard-sphere displacement
$\Delta_{\rm HS}$ is proportional to $\sin\theta/2$ and thus vanishes 
for the zero angle scattering. Similarly, the rotational displacement 
$\Delta_\phi$ is proportional to $\cos\theta/2$ so that it vanishes for 
the back scattering. The undefined directions of the displacements in 
these cases thus does not cause any problems.

\section{Numerical results}
\begin{figure}
\parbox[t]{16cm}{
\psfig{file=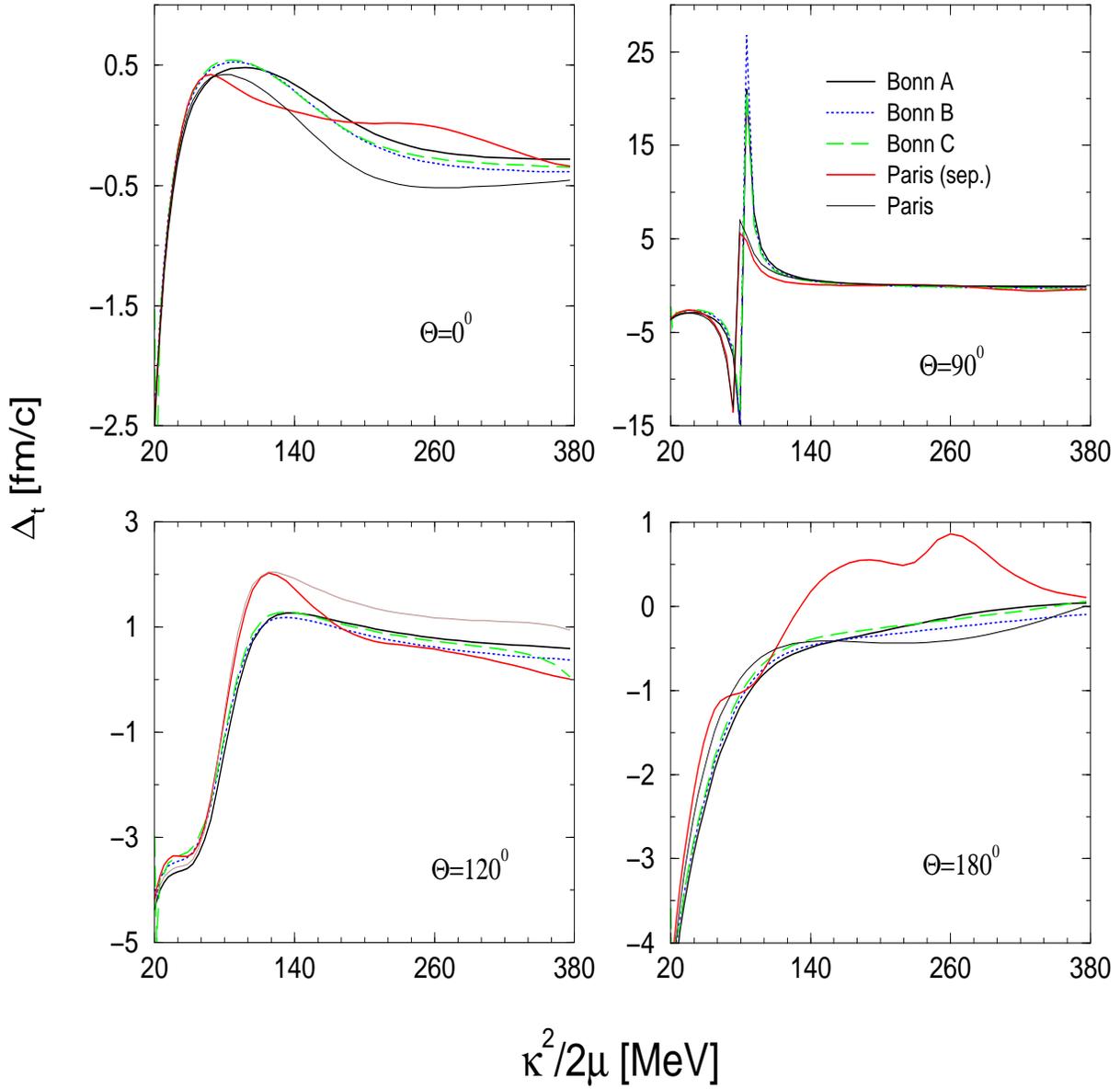,width=16cm,height=16cm,angle=-90}}
\caption{The collision duration $\Delta_t$ as a function of the 
deflection angle, $\theta$, and the kinetic energy, 
$|\kappa|^2/m\equiv\kappa^2/2\mu$, in the barycentric coordinate 
system. A dramatic change from the negative to positive collision 
duration seen for the perpendicular scattering, $\theta=90^o$, at 
energies about 80~MeV, appears for processes of a vanishing scattering 
rate. A good agreement is found between results obtained from five 
different interaction potentials, except for the back scattering, 
$\theta=180^o$, where the separable Paris potential yields appreciably 
different results from the others.
\label{dt}}
\end{figure}

\begin{figure}
\parbox[t]{16cm}{
\psfig{file=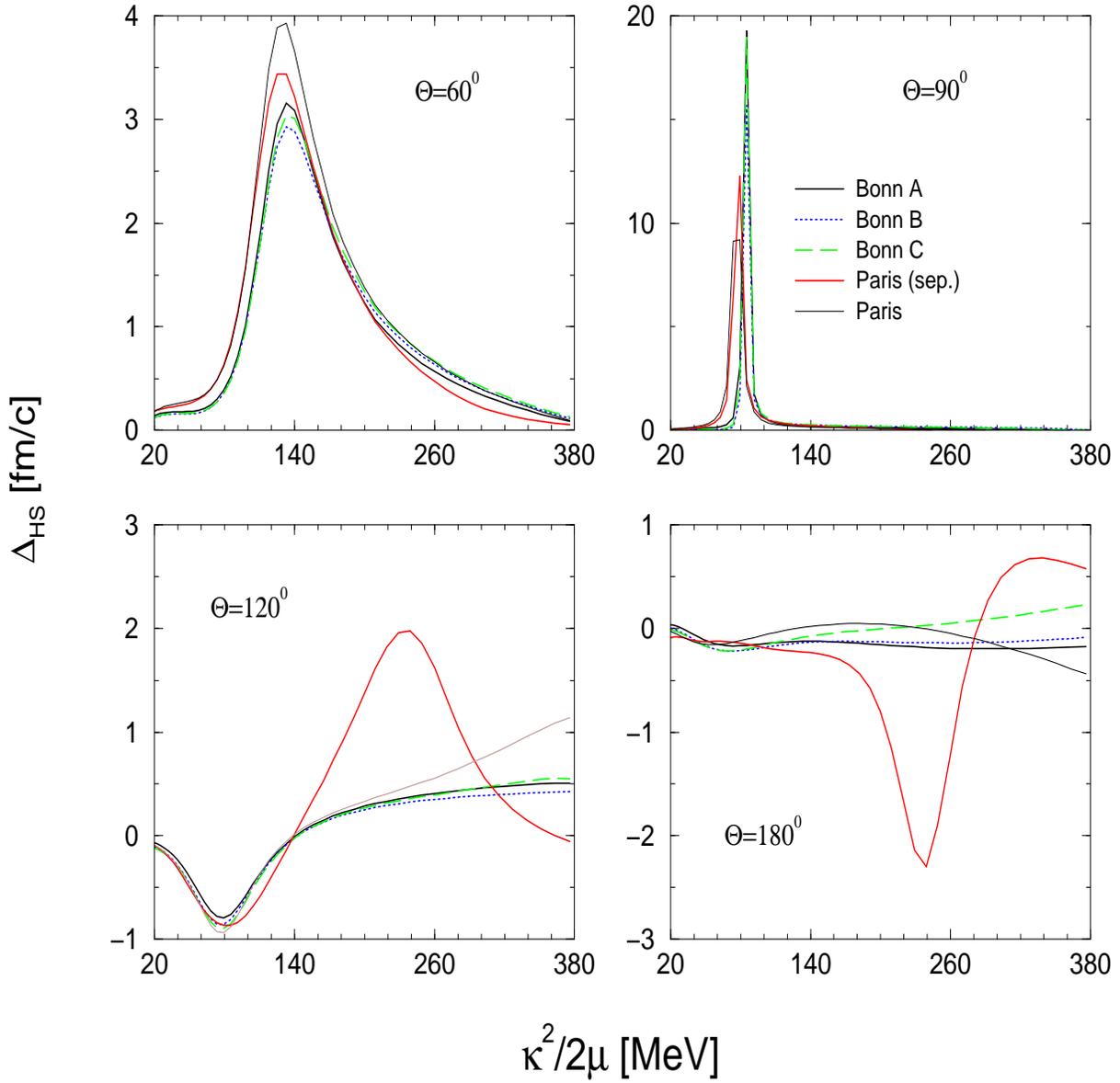,width=16cm,height=16cm,angle=-90}}
\caption{The hard-sphere displacement $\overline{\Delta_{\rm HS}}$ for 
conditions identical to Fig.~\protect\ref{dt}. The forward scattering 
is not included because $\overline{\Delta_{\rm HS}}=0$ for $\theta=0^o$.
As in Fig.~\protect\ref{dt}, processes corresponding to the singularity 
seen at $\theta=90^o$ and $\kappa^2/2\mu=80$~MeV have a vanishing 
scattering rate, and the separable approximation leads to serious 
deviations for large deflection angles, $\theta=120^o$ and 
$\theta=180^o$.}
\label{ht}
\end{figure}

\begin{figure}
\parbox[t]{16cm}{
\psfig{file=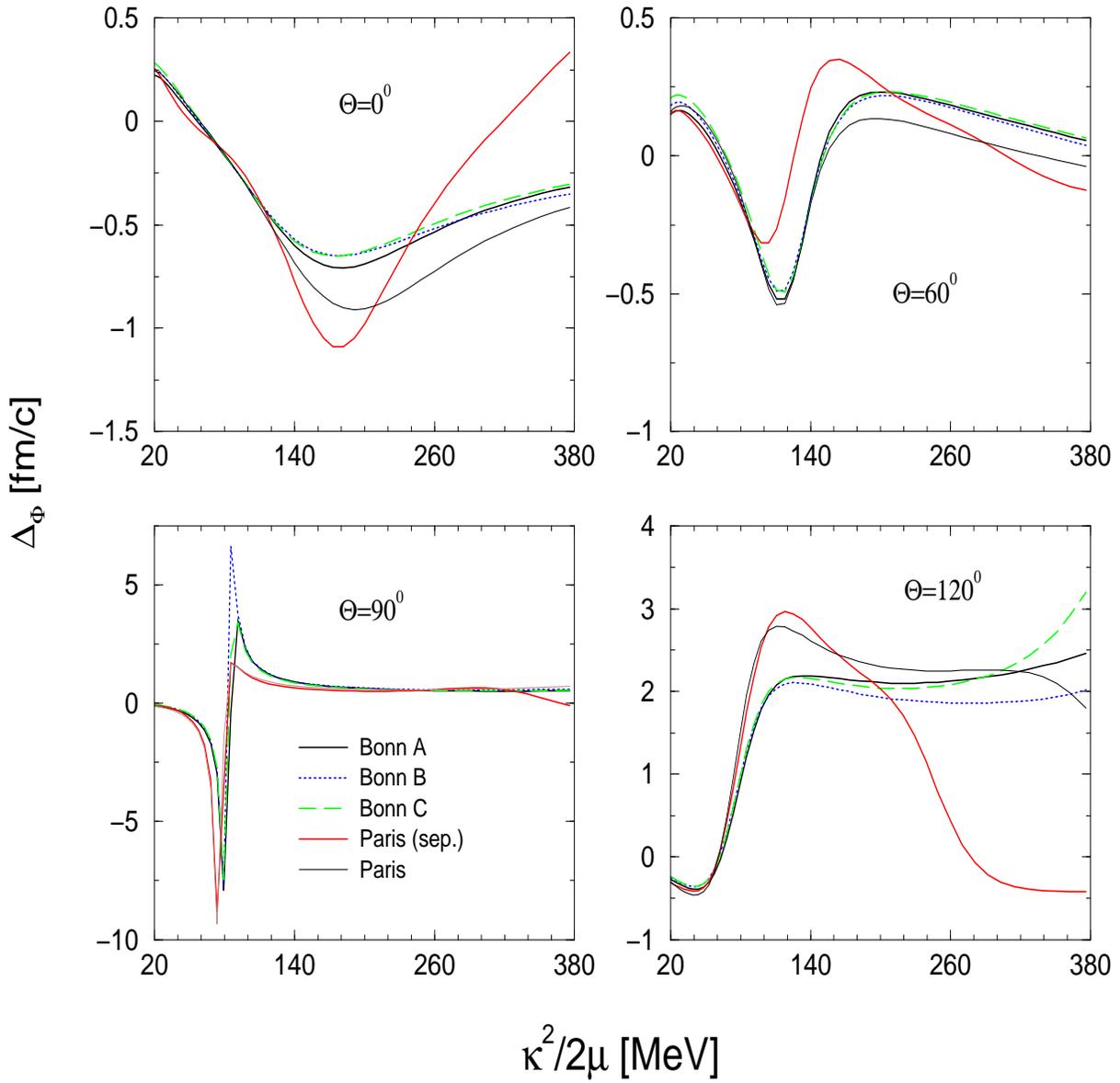,width=16cm,height=16cm,angle=-90}}
\caption{The rotational displacement $\overline{\Delta_{\phi}}$ for 
conditions identical to Fig.~\protect\ref{dt}. The back scattering 
is not included because $\overline{\Delta_\phi}=0$ for $\theta=180^o$.
Note similarities with Fig.~\protect\ref{dt} for deflection angles
$\theta=90^o$ and $\theta=120^o$.}
\label{at}
\end{figure}

Numerical values of the collision duration $\Delta_t$, the hard-sphere
displacement $\overline{\Delta_{\rm HS}}$ and the rotational displacement
$\overline{\Delta_\phi}$ are shown in Figs.~\ref{dt}-\ref{at}. They are
evaluated from (\ref{tmexp}) with the sum over partial waves terminated 
above D-waves but the coupled channels, $^3$P$_2-^3$F$_2$ and 
$^3$D$_3-^3$G$_3$, are included along with $^3$S$_1-^3$D$_1$. We compare 
results for five approximations of the nucleon-nucleon potential: 
a set (A-C) of one-Boson-exchange Bonn potentials \cite{Ma89,LMK93}, 
the Paris potential \cite{LLRVCPT80} and the separable Paris potential 
\cite{HP84}.

\subsection{Collision duration}
In figure \ref{dt} we plot the collision duration for different scattering
angles versus lab energy. Three features are apparent: (i) at low energies 
the collision duration reaches large negative values for all deflection 
angles, (ii) at higher energies the collision duration strongly depends on 
the deflection angle, and (iii) sharp discontinuities, as the one seen for
$\theta=90^o$ at energy 80~MeV, might appear.

The large negative values (i) of the collision duration at low energies
follow from general properties of the T-matrix. The real part, Re$T$, is 
regular for $\Omega\to 0$ while the energy dependence of the imaginary 
part, Im$T$, is proportional to the density of states, Im$T\propto
\sqrt{\Omega}$. At low energies, the collision duration hence behaves as 
$\Delta_t\approx{1\over{\rm Re}T}{\partial{\rm Im}T\over\partial\Omega}
\propto{1\over\sqrt{\Omega}}$. Using the on-shell condition, $\Omega=
{|\kappa|^2\over m}$, one can express this singularity as 
$\Delta_t\propto{1\over|\kappa|}$. Such a singularity is not dangerous
for the kinetic equation since the mean time between collisions also
scales with ${1\over|\kappa|}$ being inversely proportional to velocity.

The negative value of the collision duration shows that positions of 
nucleons are anti-correlated, i.e., it is less likely to find them in 
a close vicinity than it would be the case in the absence of the 
interaction. This feature appears also in classical systems, where two 
particles speed up their motion in the range of their attractive 
potential passing each other faster than in the absence of forces. 

The angular dependence (ii) of the collision duration is surprisingly
irregular. The back scattering, $\theta=180^o$, reveals monotonic energy 
dependence of the collision duration. The forward scattering, $\theta=0^o$, 
has a single maximum at 100~MeV. The perpendicular scattering, 
$\theta=90^o$, has a ${1\over x}$-singularity at 80~MeV. Finally, the 
scattering at $\theta=120^o$ has a step at 30~MeV which likely results 
from a smoothened ${1\over x}$-singularity at 80~MeV. 

The strong angular dependence shows that an interference between S, P, D 
waves plays the important role in the collision duration. For instance, 
at $\theta=0^o$ all Legendre polynomials equal to unity, $P_{0,1,2}=1$, 
therefore the T-matrix reads $T={1\over 16\pi}(T_0+T_1+T_2)$. At 
$\theta=180^o$, $P_{0,2}=1$ and $P_1=-1$, therefore $T={1\over 16\pi}
(T_0-T_1+T_2)$. The difference between the collision duration at 
$\theta=0^o$ and $\theta=180^o$ thus follows from a different 
interference of the P wave with the other waves.

The angular dependence reduces at energies below 30~MeV. For this energy,
the de Broglie wave length ${\hbar\over|\kappa|}\sim1$~fm is comparable 
to the range of the interaction potential. With increasing wave length, the 
interaction potential behaves more and more as a contact potential, 
therefore the S component of the scattered wave dominates at all angles.

In the ${1\over x}$-discontinuity (iii) at 80~MeV and $\theta=90^o$, the
collision duration reaches so high values that one might wonder why the
instantaneous approximation of the collision works so well. In fact, this
type of singularity is not dangerous for kinetic equations, because it 
appears when the differential cross section vanishes. To see it in more
detail, let us write the T-matrix. For $\theta=90^o$ we have $P_0=1$,
$P_1=0$ and $P_2=-{1\over 2}$, therefore $T={1\over 16\pi}(T_0-{1\over 2}
T_2)$. At the energy of 80~MeV the S and D waves destructively interfere
and $T\to 0$ while its energy derivative remains finite, ${\partial T
\over\partial\Omega}\not=0$. This causes the ${1\over x}$-discontinuity
seen in $\Delta_t$. In the kinetic equation the collision duration 
enters in a product with the cross section, as a gradient contribution 
proportional to $|T|^2\Delta_t$. At the point of singularity, 
$|T|^2\Delta_t\to 0$. It is possible that the ${1\over x}$-discontinuity 
will vanish or get reduced when more partial waves will be included. 

Let us compare the collision delays found from different potentials.
All members from the set of Bonn potentials provide nearly identical 
results. There is also only a minor difference between the Paris and the 
Bonn potentials. The strongest difference appears between the separable 
Paris potential and the other for the back scattering. Nevertheless,
one can conclude that on the scale of precision of recent theory of
heavy ion reactions, all potentials are equally acceptable.

\subsection{Hard-sphere displacement}
The hard-sphere displacement $\overline{\Delta_{\rm HS}}$ is plotted in 
Fig.~\ref{ht}. Unlike the collision duration, the displacement is regular
at low energies going to zero for $|\kappa|\to 0$. This is caused by the
dominant contribution of the S wave in this region. When the S wave
dominates, the hard-sphere displacement simplifies as
$\overline{\Delta_{\rm HS}}=-{1\over 2}\sin{\theta\over 2}\,{\rm Im}
{1\over T_0}{\partial T_0\over\partial|\kappa|}$. From analyticity 
follows that at small momenta the T-matrix depends on $|\kappa|^2$, 
therefore ${\partial T_0\over\partial|\kappa|}\propto|\kappa|$ and 
vanishes for $|\kappa|\to 0$.

Again, there is a strong and irregular angular dependence. The angular 
variation of $\overline{\Delta_{\rm HS}}$, visible for deflection angles
$\theta=60^o,90^o,120^o$, clearly shows that the model of hard spheres
is not adequate for any energy region, because it yields the 
angle-independent length of displacement. It is interesting that for the 
non-separable potentials the effective hard sphere is very small at the 
back scattering, $\theta=180^o$, in spite of the maximum of the factor 
$\sin\theta/2$. The negative value of the length 
$\overline{\Delta_{\rm HS}}$ at $\theta=120^o$ shows that the non-local 
corrections to the nucleon-nucleon collision cannot be estimated from 
the hard core of the nucleon-nucleon potential. If the effect of the 
real hard core would dominates the non-locality, the length of 
displacement is expected to be positive and equal twice the radius of 
the core.

Finally, the perpendicular scattering has a singularity at energy 80~MeV.
Similarly to the singularity of the collision duration, the enormous
values of non-local corrections at this point are not dangerous for the 
validity of the kinetic equation, because they appear due to vanishing 
differential cross sections. Traces of this singularity can be seen also
for scattering at close angles, $\theta=60^o$ and $\theta=120^o$.

The non-separable potentials yield quite close values of the hard-sphere
displacement. Certain difference between the Bonn and Paris potentials 
appears at low energies for $\theta=60^o$. Recent precision of 
heavy-ion-reaction simulations, however, does not allow 
to distinguish such a detail. Strong differences result only between
the separable and non-separable potentials, in particular, for the back
scattering.

\subsection{Rotational displacement}
The rotational displacement $\overline{\Delta_\phi}$ is plotted in 
Fig~\ref{at}. In the limit of low energies it vanishes from a similar 
reason as the hard-sphere displacement. When the S wave dominates, the 
rotational displacement simplifies as $\overline{\Delta_\phi}={1\over 2}
\cos{\theta\over 2}\,{\rm Im}{1\over T_0}{\partial T_0\over\partial
|\kappa|}$ which goes to zero as $|\kappa|\to 0$.

Except for the low energy region, the rotation displacement has a number
of features similar to the collision duration, at least for $\theta=90^o$ 
and $\theta=120^o$. For the perpendicular scattering, the 
${1\over x}$-discontinuity appears at 80~MeV. The step for $\theta=120^o$
at 40~MeV is also similar including the shoulder at 120~MeV. The velocity
of particles in the barycentric framework for 80~MeV is about a half of
the velocity of light, which crudely corresponds to the coefficient with
which the discontinuity of the collision duration scales on the 
discontinuity of the rotational displacement. We are not aware of the 
real reason for this similarity, nevertheless, it is noteworthy that for 
a collision of classical particles the rotational displacement is also
proportional to the collision duration, because the longer the collision 
lasts the more the colliding pair can revolve. For instantaneous 
collisions, no rotational displacement appears due to a continuity of 
classical trajectories, see Fig.~\ref{soft}. In contrast, the hard-sphere 
displacement is generally non-zero for instantaneous collisions as it is 
the case for the true hard spheres. Small correlation between $\Delta_t$
and $\Delta_{\rm HS}$ is also confirmed by the numerical results, at 
least figures~\ref{dt} and \ref{ht} do not show similar features.

Except for singularities, which are invisible due to the small 
differential cross section, the typical values of the rotational
displacement are slightly smaller but close to the hard-sphere
displacement. In studies \cite{M84,KDB96,DP96} the rotational 
displacement has not been assumed. The found numerical values do not 
justify this neglect.

\subsection{Wigner collision delay}
At the low energy region, ${|\kappa|^2\over m}<30$~MeV, the collision
duration becomes a sizable correction while the displacements are rather 
small. This energy domain is important for heavy ion collisions at 
medium energies, it is thus worthy to test the validity of the 
approximation by the Wigner collision delay in more details. 

To make a comparison of the non-instantaneous and the non-local 
corrections easier, we express both in the form of displacements which 
enter the simulation codes, as it has been done in \cite{MLSCN98}. 
Assuming that the collision event is identified at the instant of 
the closest approach of particles, the total effective displacement in 
the direction of outgoing relative momentum, $\Delta_\|={\kappa_{\rm f}
\over|\kappa|}\overline{\Delta_\|}$, has a length
\begin{eqnarray}
\overline{\Delta_\|}&=&
{\rm Im}\left[{1\over T}\sum\limits_lP_l\left(
{\partial T_l\over\partial\Omega}{2|\kappa|\over m}+
{\partial T_l\over\partial|\kappa|}\right)\right]
\nonumber\\
&=&{2|\kappa|\over m}
{\rm Im}\left[{1\over T}\sum\limits_lP_l\left(
{\partial T_l\over\partial\Omega}+
{\partial T_l\over\partial{|\kappa|^2\over m}}\right)\right]
\nonumber\\
&=&{2|\kappa|\over m}\Delta_t^{\rm W},
\label{wigdel}
\end{eqnarray}
This displacement combines the collision duration, the final relative
displacement ${1\over 2}(\Delta_4-\Delta_3)=\Delta_{\rm HS}+\Delta_\phi$
projected on the direction of the final momentum $\kappa_{\rm f}$, and
the initial relative displacement ${1\over 2}\Delta_2=\Delta_{\rm HS}-
\Delta_\phi$ projected on the direction of the initial momentum $\kappa$.
For details of the instantaneous approximation see \cite{MLSCN98}. The 
same displacement results when one derives the non-local correction 
from the Wigner collision delay $\Delta_t^{\rm W}$. 

The displacement perpendicular (inside the collision plane) to the 
direction of the outgoing relative momentum has a length
\begin{equation}
\overline{\Delta_\perp}=-{\sin\theta\over|\kappa|}\ {\rm Im}\left[
{1\over T}\sum\limits_lP'_lT_l\right].
\label{delperp}
\end{equation}
This perpendicular component is neglected within the approximation
derived from the Winger collision delay.

\begin{figure}
\begin{minipage}{7cm}
\parbox[t]{7cm}{
  \psfig{figure=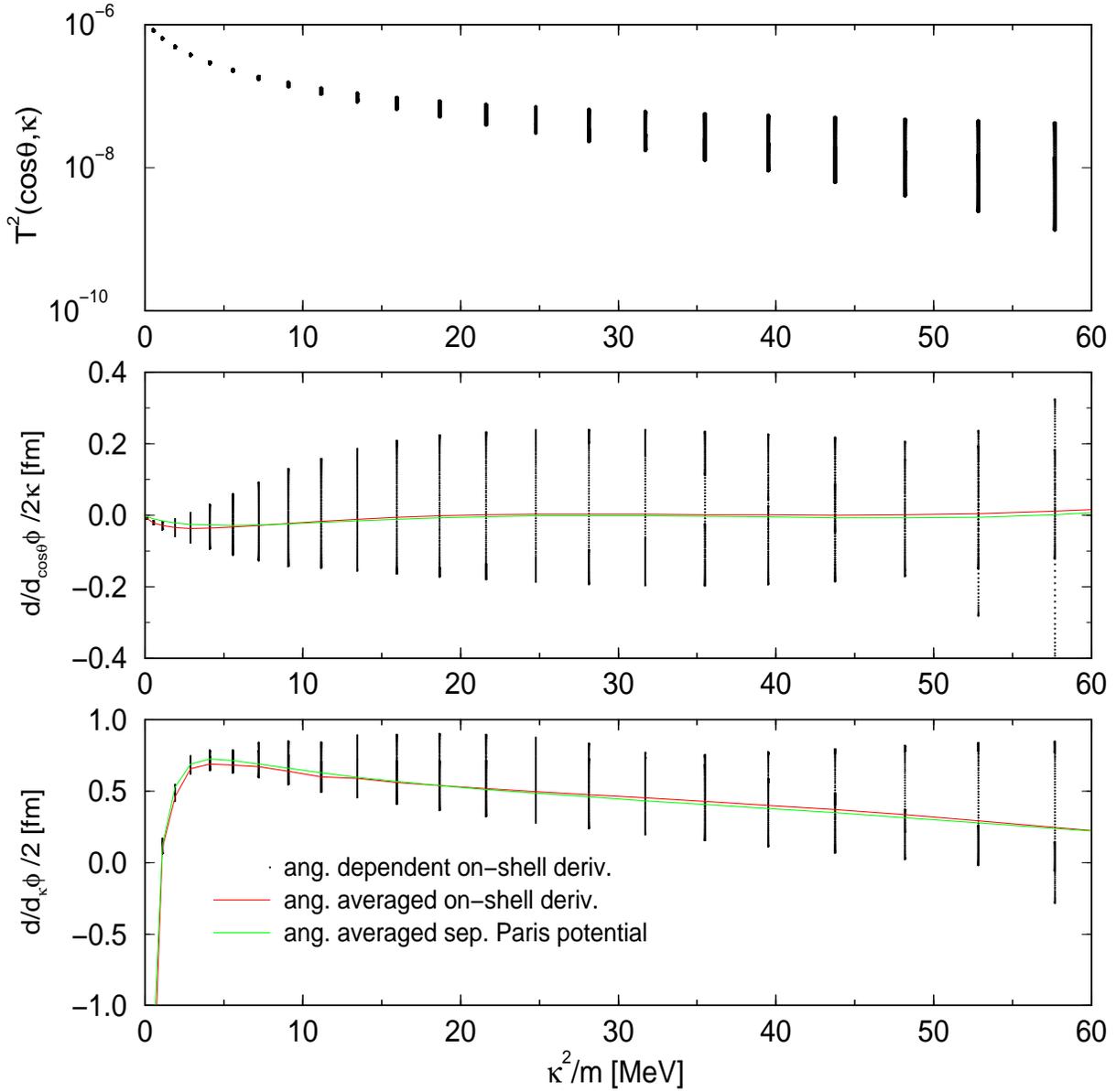,width=16cm,height=16cm,angle=-90}}
\end{minipage}
\caption{The effective displacement as a function of the deflection
angle and the kinetic energy, $\kappa^2/m$, in the barycentric
coordinate system. The columns of dots show the spread of components 
with deflection angle. The lines show the angle-averaged values. The 
amplitude of the T-matrix is presented in the top section to indicate 
the weight of individual processes. The orthogonal component, 
$\overline{\Delta_\perp}\equiv d/d_{\cos\theta}\phi$ shown in the 
middle section, has appreciably smaller values than the parallel 
component, $\overline{\Delta_\|}\equiv d/d_\kappa\phi/2$ shown in 
the bottom section.}
\label{ixf1}
\end{figure}

Numerical values of these two contributions are compared in 
Fig.~\ref{ixf1}. The dots in the vertical line show a spread of values 
due to the angular dependence, the curves show values averaged over 
deflection angles with the weight given by the differential cross 
section displayed in the top section. The parallel component, shown in 
the bottom section, has a typical value of 0.5~fm. The negative large 
values below 3~MeV can be ignored since corresponding processes have 
very small rates due to the Pauli blocking. The perpendicular component,
shown in the middle section, has about three-times smaller values, 
moreover it tends to average out. These results confirm that the
approximation based on the Wigner collision delay as proposed by 
Danielewicz and Pratt \cite{DP96} is suitable for the nuclear matter 
at low energies.

\section{Summary}
The numerical values of the collision duration and displacements of
particles in a binary collision are comparable with typical time and 
space scales in reacting heavy ions. This shows that non-instantaneous 
and non-local treatments of binary collisions in simulations of heavy 
ion reactions are desirable. 

The found collision duration and displacements slightly depend on the
employed interaction potential. The differences are, however, too small
to be detectable within the accuracy of recent realistic simulations. 
Stronger deviations have 
appeared for the separable potential at higher energies. This was 
expected, since the separable approximation is suited for the low 
energy region only.

We have tested the approximations of non-local corrections adopted or 
suggested in print. The energy and angular dependence of the duration 
and displacements do not justify the hard-sphere model. On the other 
hand, the approximation by the Wigner collision delay covers the 
dominant non-local corrections in the low energy region. 

\acknowledgements
The authors are grateful to S. K{\"o}hler for stimulating discussions.
This work was supported from the Czech republic, GACR Nos.~202960098 
and 202960021 and GAASCR Nr.~A1010806, and Germany, BMBF Nr.~06R0884,
the Max-Planck Society.

\vspace{3ex}

\onecolumn

\end{document}